\documentclass[preprint,showpacs,preprintnumbers,amsmath,amssymb]{revtex4}

\usepackage{epsfig}
\usepackage{dcolumn}
\usepackage{bm}

\def\e{\begin{equation}}
\def\f{\end{equation}}
\def\%#1{\mbox{\boldmath $#1$}}
\def\=#1{\overline{\overline #1}}

\def\_#1{{\bf #1}}

\def\E{\varepsilon}

\def\.{\cdot}
\def\x{\times}
\def\##1{{\bf#1\mit}}

\def\l#1{\label{eq:#1}}
\def\r#1{(\ref{eq:#1})}
\def\am{\left(\begin{array}{c}}
\def\amm{\left(\begin{array}{cc}}
\def\a{\end{array}\right)}

\begin{document}


\title{Sub-wavelength imaging: Resolution enhancement using metal wire gratings}

\author{G. Fedorov$^1$, S.I. Maslovski$^2$, A.V. Dorofeenko$^1$,  A.P. Vinogradov$^1$, I.A. Ryzhikov$^1$,
S.A. Tretyakov$^2$}%
 \affiliation{$^1$Scientific Center for Applied Problems in Electrodynamics\\
 Russian Academy of Sciencies\\
 Izhorskaya 13/19, Moscow\\
 127412 IVTAN, Russian Federation}
\affiliation{$^2$Radio Laboratory/SMARAD\\
Helsinki University of Technology\\
P.O. Box 3000, FI-02015 TKK, Finland}

\date{\today}

\begin{abstract}

An experimental evidence of subwavelength imaging with a ``lens",
which is a uniaxial negative permittivity wire medium slab, is
reported. The slab is formed by gratings of long thin parallel conducting
cylinders. Taking into account the
anisotropy and spatial dispersion in the wire medium we
theoretically show that there are no usual plasmons that could be
exited on surfaces of such a slab, and there is no resonant
enhancement of evanescent fields in the slab. The
experimentally observed clear improvement of the resolution in the
presence of the slab is explained as filtering out the harmonics
with small wavenumbers. In other words, the wire gratings (the
wire medium) suppress strong traveling-mode components increasing
the role of evanescent waves in the image formation.  This
effect can be used in near-field imaging and detection
applications.
\end{abstract}

\pacs{71.45, 73.20.Mf, 41.20, 42.25}
\keywords{surface plasmon, wire medium, near field, resolution, slab lens}
\maketitle

\section{Introduction}

Recently, much effort has been devoted to studies of novel
sub-wavelength focusing and imaging systems. Sir John Pendry was
the first to suggest employing slabs of materials with negative
permittivity and permeability to overcome the
Rayleigh limit in imaging \cite{Pendry}. Since then there have been published
many papers confirming Pendry's prediction both in computer
simulations and experimentally \cite{exp}. Such imaging becomes possible
due to resonant excitation of surface plasmons on the slab
surfaces by evanescent waves existing near an object. These
surface plasmons subsequently excite evanescent waves reproducing
the object's near fields in the image plane. In other words, the
super-resolution is achieved by amplification of near fields by
plasmon resonances (e.g., \cite{maslovski}). The known realization of the effect in the
microwave region is based on the use of a composite material that
combines an array of thin conducting wires (wire medium) and an
array of split rings. Due to the inductances of wires forming the
medium, the wire medium behaves like an anisotropic material with
negative permittivity for the electric field component
parallel to the wires and positive (close to unity) permittivity
for the electric field component perpendicular to the wires (e.g.
\cite{Brown,Rotman,Belov,modeboo}). Near the fundamental resonance of the currents induced in
split rings, the induced magnetic moments can be out of phase with
the incident magnetic field, realizing negative effective
permeability \cite{sch,pendry_srr}. In accordance with the theory \cite{Pendry}, a slab of
a material with only negative permittivity can also be used for
focusing and imaging waves of the TM polarization. There are many
materials exhibiting negative permittivity in optics, but at
microwaves there are no low-loss natural materials that would
provide negative real part of permittivity. To realize the effect
at microwaves, one might try to use the same wire medium as a
negative-epsilon metamaterial \cite{Ramakrishna}, as in the known experiments
with double-negative metamaterials. The use of only wire medium could
have advantages of a wide-band frequency response, because there
are no resonant split rings. Actually even single or double wire
gratings could be used for this purpose.

In this paper we provide
experimental data on imaging by means of wire media and explain
the phenomena observed in our experiment. The experiment shows
that a dense wire grating or a pair of such gratings put in an
arrangement similar to that of \cite{Lagarkov} can provide subwavelength
imaging. We use dipole antennas as the field sources and
detectors. In our experiment the radiated by antenna field is
rather of S-polarization [transversal-electric (TE)], than of
P-polarization [transversal-magnetic (TM)]. According to \cite{Pendry},
only fields of TM-polarization take part in subwavelength imaging
by a negative-epsilon slab. Thus, the experiment is not a simple
confirmation of the theory \cite{Pendry} but shows new phenomena. Since it
is known that subwavelength imaging is due to excitation of
surface waves (plasmons) we study carefully what kind of surface
waves can exist on an interface between free space and a wire
medium formed by infinite ideally conducting cylinders (wires)
placed in free space. The wires are assumed to be thin so that the
relative transversal dielectric permittivity is close to unity.
The interface is parallel to the wires.

We show that when the spatial dispersion in wire medium \cite{Belov_disp} is properly
taken into
account, there are no usual surface plasmons of TM polarization.
Instead, there are waves that behave like modes of a multi-wire
transmission line. In a
multi-wire transmission line composed of $N$ wires there are exactly $N$
modes. Although
these modes differ in the distribution of currents among the wires, they
all propagate with
the speed of light along the wires. In a system of infinite number of
wires (wire medium)
there is an infinite number of similar transmission-line modes.
Superimposing these modes one can get an
arbitrary distribution of wire currents. On the macroscopic scale (on
the whole medium
scale) that means that for such modes the wave vector component
orthogonal to the wires
can be arbitrary, however, the component along the wires always equals
$\omega/c$. We show that
unlike usual plasmons in metals, these modes of a uniaxial wire medium
do not exhibit
a resonance interaction with incident evanescent fields.
Therefore, there is no resonant enhancement of evanescent
fields performed by a slab of such medium. However, our
experiments show a clear improvement of the resolution in the
presence of wire medium slabs. We explain this as filtering out
the harmonics with small wavenumbers performed by wire gratings of
which the slabs are composed. In other words, wire gratings
suppress strong traveling-mode components increasing the role of
evanescent waves in the formation of the image.

\section{Experiment}

\begin{figure}[htb]
\centering \epsfig{file=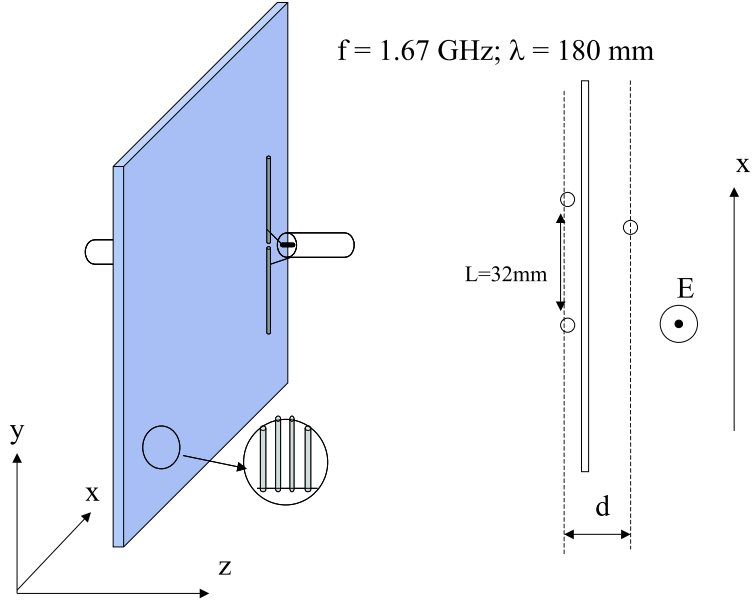, width=0.6\textwidth}
\caption{Experimental configuration.} \label{fig1}
\end{figure}

\begin{figure}[ht]
\centering \epsfig{file=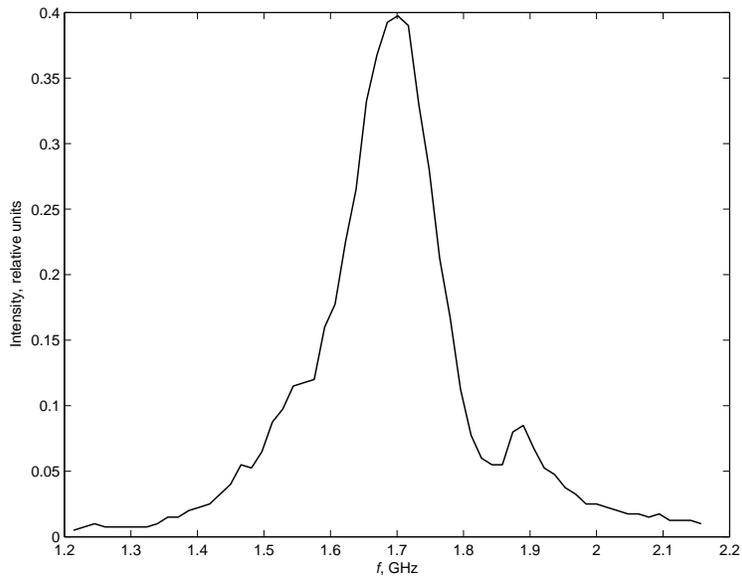, width=0.6\textwidth}
\caption{Signal in the probe antenna as a function of frequency.}
\label{fig2}
\end{figure}

The experimental set-up is shown in Fig.~\ref{fig1}. One or two
planar wire grids formed by round copper wires of 8 microns in
diameter and 30 cm in length were excited by two parallel
half-wavelength dipoles. The source dipoles were parallel to the
wires (this arrangement is similar to that of Lagarkov et al.
\cite{Lagarkov}, who used an artificial double-negative slab
formed by wires and split-ring resonators). The grid period
(distance between the wires) was 2 mm. The values of the other
quantities are shown in Fig.~\ref{fig1}. The working frequency
($f=1.67$ GHz) was determined experimentally as the one that
provided the maximum signal in the probe antenna in the image
plane in the absence of wire grids (Fig.~\ref{fig2}).

\begin{figure}[ht]
\centering \epsfig{file=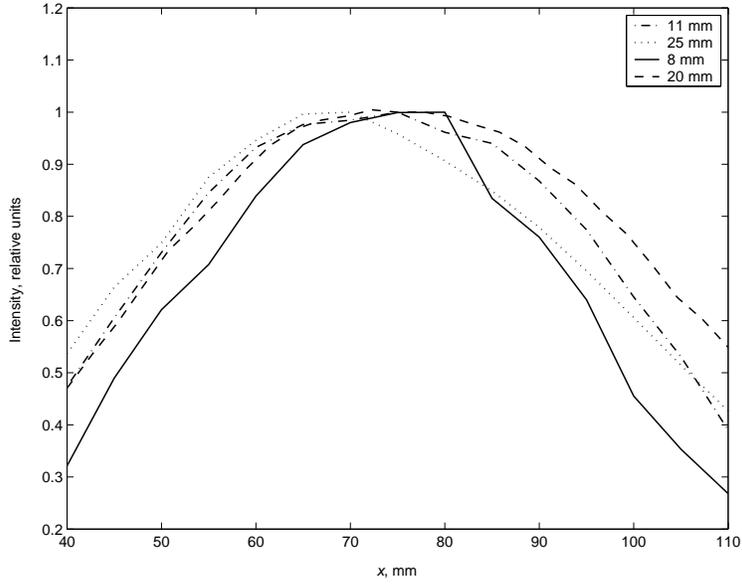, width=0.6\textwidth}
\caption{The measured intensity as a function of the position. The
grid is absent. Different curves correspond to different distances
$d$ from the grid to the receiving antenna.} \label{fig3}
\end{figure}

\begin{figure}[ht]
\centering \epsfig{file=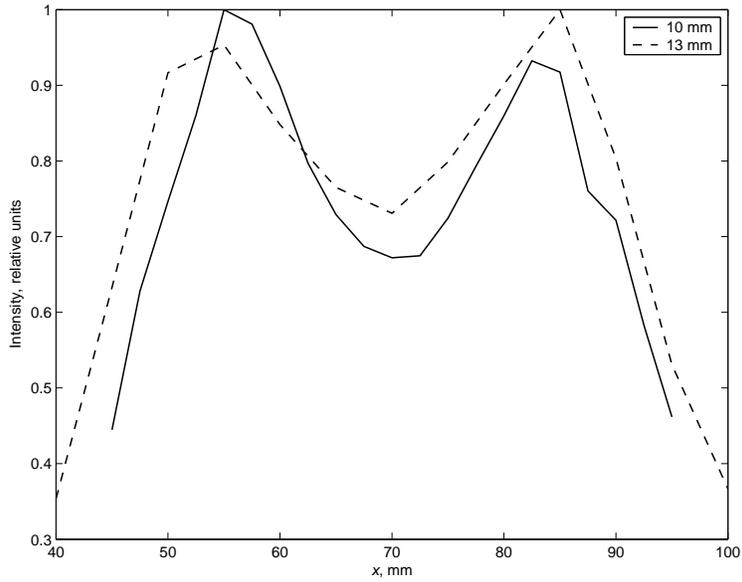, width=0.6\textwidth}
\caption{The same as in Fig.~\ref{fig3} but with one grid present.}
\label{fig4}
\end{figure}

\begin{figure}[ht]
\centering \epsfig{file=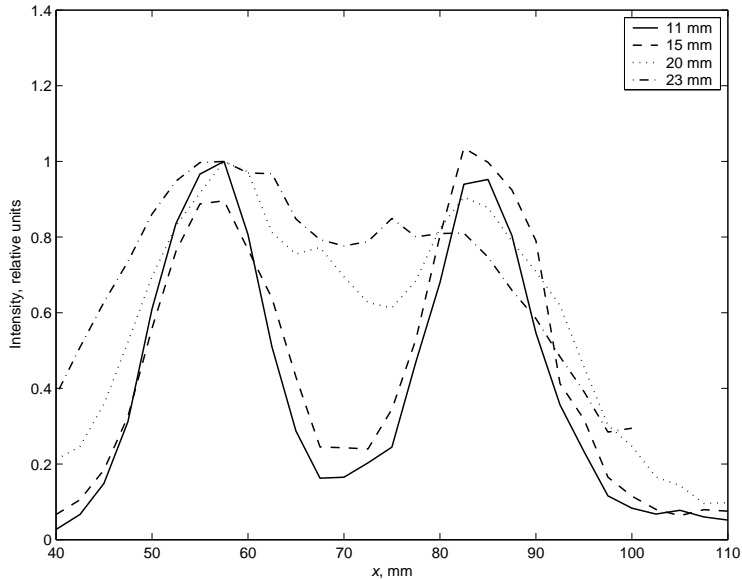, width=0.6\textwidth}
\caption{The same as in Fig.~\ref{fig3} but with two grids present.}
\label{fig5}
\end{figure}

In Figs.~\ref{fig3}--\ref{fig5} the intensity measured by the
probe antenna behind the grid as a function of the position along
the axis $x$ is shown. We can see that without the grid we have a
single maximum, whereas in the system with one or two grids we can
see two maxima.

In the near-field zone, the electric fields of the two antennas
have non-zero components along any direction. The electric field
component along the wires excites the grid currents. Because the
distance between the source dipoles is about $\lambda/6$, the
experimental results show that the device is capable to resolve
sub-wavelength details.

Next, we will prove that strong spatial dispersion in uniaxial wire medium
supresses resonant excitation of surface plasmons. An alternative
explanation of the observed subwavelength resolution will be found.

\section{Transmission line modes instead of plasmons}

We will look for plasmons on the surface of wire medium formed by thin parallel
ideally conducting cylinders densely (in terms of the wavelength) packed in
free space. This artificial material can be modeled by a uniaxial
relative permittivity dyadic whose longitudinal (along the wire
direction) component depends on the frequency and the longitudinal
wavenumber \cite{Belov_disp}
\begin{equation}
\=\varepsilon_r = \E_\parallel\_u_0\_u_0 + \E_\perp(\=I - \_u_0\_u_0),
\label{eq:1}
\end{equation}
where $\_u_0$ is the unit vector along the wires, and $\=I$ is the
unit dyadic. The components of the dielectric permittivity dyadic are
$\E_\parallel$ and $ \E_\perp$ (for thin wires):
\begin{equation}
\varepsilon_{\parallel}=\left(1-{k_p^2\over{k_0^2-k_\parallel^2}}\right), \qquad \varepsilon_\perp\approx 1.
\l{permit}
\end{equation}
Here $k_p$ is the plasma wavenumber (e.g., \cite{Belov_disp}),
$k_\parallel$ is the propagation constant component along the
wires, and $k_0$ is the free-space wavenumber.

If the interface between the wire medium (seen as a composition of many
parallel ideally conducting cylindtical wires) and free space is parallel to the direction of the wires,
the whole system is regular along that direction. This means that all the eigenmodes of
such a system (including surface modes) split into ordinary and
extraordinary modes, see, for example \cite{modeboo}. The ordinary modes
have zero electric field component along the wires (TE modes with
respect to the direction along the wires), and the
extraordinary modes have the corresponding magnetic field
component equal to zero (TM modes with respect to the wires).
Obviously, the ordinary (TE) modes are the same as
the waves in free space because they are not affected by
thin wires. Therefore, there are no surface
modes (surface plasmons) among the ordinary (TE) eigenmodes.
There are also modes with both electric and magnetic field
components vanishing along the wires (TEM modes with respect to the
wires). We may consider them as a limiting case of the extraordinary
(TM) modes.

To find the dispersion relation for the surface plasmons of
extraordinary polarization one has to equate the tangential
electric and magnetic fields at the interface of the wire medium
and free space. We denote $E_\parallel = \_u_0\.\_E$ and $H_\perp
= (\_u_0\x\_n)\.\_H$, where $\_n$ is the unit normal. In these
notations $H_\parallel = 0$ (we are interested only in
extraordinary modes). The same notation is used for the components
of the wave vector $\_k$. Then the following relation between
$E_\parallel$ and $H_\perp$ can be derived from the Maxwell
equations:
\begin{equation}
{E_\parallel\over H_\perp} = \pm \sqrt{\mu_0\over\E_0}{k_0^2 -
k_\parallel^2\over k_0\left[(k_0^2 - k_\parallel^2)\E_\parallel -
  k_\perp^2\right]^{1\over 2}}=
\pm \sqrt{\mu_0\over\E_0}{k_0^2 -
k_\parallel^2\over k_0\left[k_0^2 - k_p^2 - k_\perp^2 -
  k_\parallel^2\right]^{1\over 2}}\,.
\l{wmimp}
\end{equation}
Different signs in this formula correspond to two possible directions
of propagation (decay) with respect to $\_n$. The square root in the
denominator of this formula is the magnitude of the normal component
of the wave vector. When dealing with surface modes, the argument of
the square root is negative and the normal component of the wavevector
is imaginary. In this case the following branch of the square root
must be used:
\begin{equation}
\Big[k_0^2 - k_p^2 - k_\perp^2 - k_\parallel^2 \Big]^{1\over 2} =
-j\sqrt{k_p^2 + k_\perp^2 + k_\parallel^2 - k_0^2}.
\end{equation}

Here we give also the dispersion equation of the extraordinary
modes in an {\em unbounded} wire medium. This is obtained in the
same derivation procedure along with \r{wmimp}: \e k^2 = k_0^2 -
k_p^2. \l{dispmed} \f

The waves in the free space region can be also formally separated into ordinary and extraordinary modes of the same definition. Then, for the extraordinary modes we obtain
\begin{equation}
{E_\parallel\over H_\perp} = \pm \sqrt{\mu_0\over\E_0}{k_0^2 -
k_\parallel^2\over k_0\left[k_0^2 - k_\perp^2 -
  k_\parallel^2\right]^{1\over 2}}\,.
\l{fsimp}
\end{equation}
In general case $E_\perp\neq 0$. The following relation holds for an
extraordinary (TM) wave
\begin{equation}
{E_\perp\over E_\parallel} =
-{k_\parallel k_\perp\over k_0^2 - k_\parallel^2}\,.
\l{cross}
\end{equation}
This relation is the same in the wire medium  and in free-space (in the
case when $\E_\perp = 1$) and, therefore, it is not necessary for the
derivation of the dispersion relation.

The dispersion relation of the  surface plasmons is obtained by
equating \r{wmimp} and \r{fsimp} for the modes both decaying {\em from}
the interface: \e (k_0^2 - k_\parallel^2)\left(\sqrt{k_p^2 +
k_\perp^2 + k_\parallel^2  - k_0^2} + \sqrt{k_\perp^2 +
k_\parallel^2 - k_0^2}\right) = 0. \l{mydisp} \f The arguments of
these two square roots are positive (we are interested in
surface waves), that means that the only possible solution is
$k_\parallel^2 = k_0^2$.

The perpendicular wave vector component $k_\perp$ disappears  from
the dispersion equation, which means that it can be {\em
arbitrary}. When $k_\perp = 0$ this solution corresponds to a
simple plane wave propagating along the wires with the speed of
light. It does not decay from the interface, therefore it is not a
surface wave. However, when $k_\perp \ne 0$ the wave propagation
factor in the interface plane is $\sqrt{k_\parallel^2 + k_\perp^2}
> k_0$ which gives a wave bounded to the surface both in free
space and in the medium. Indeed, in the medium, using dispersion
relation \r{dispmed} we have for the normal component of the wave
vector (remember that $k_\parallel^2 = k_0^2$): \e k_n^2 = k^2 -
k_\parallel^2 - k_\perp^2 = -k_p^2 - k_\perp^2. \f

Perhaps, it is worth noting that this specific solution that we
have just discussed is not a surface plasmon in the usual meaning.
Instead, it results from the fact that a systems of many parallel
wires can support transmission-line type modes (TL modes). In our
case they appear as the limit of the extraordinary (TM) modes (which
have $H_\parallel = 0$) when $k_\parallel\rightarrow\pm k_0$. Indeed, from
\r{fsimp} it is seen that also $E_\parallel\rightarrow 0$, and we obtain a
TL mode as a limit of a TM mode. TL modes of a multi-wire
transmission line may have arbitrary phases of separate wire currents
which corresponds to the arbitrary $k_\perp$ in our case of the wire crystal. Still, the
propagation factor along the wires is the same, does not depend on
$k_\perp$, and equals $k_0$.

\section{Filtering instead of a resonance}
In this section we will show that the amplitude of excitation of
the TL modes considered above remains finite in structures
with infinitely long wires. This is because the incident field spectral components
which are in phase synchronism with the TL modes are also TEM
with respect to the wires. The electric field of such components
vanishes along the wires. We will show that the susceptibility of
a wire grating to an external electric field grows with the same
proportion as the longitudinal electric field decays when an
incident wave transforms from TM to TEM. Because of this the
induced grating currents remain finite and do not exhibit a
resonance.

Diffraction of plane waves by wire gratings has been well studied
in the literature. We will use the formulas from
\cite{Yatsenko,modeboo} which are applicable to thin wires. The
current $I_{\rm w}$ induced in a chosen reference wire of a
grating can be expressed in terms of the external field complex
amplitude on the wire axis $E_\parallel$, the frequency, the wave
vector, and the parameters of the grating:
\begin{eqnarray}
I_{\rm w}(k_\perp,k_\parallel) = Y(k_\perp,k_\parallel)E_\parallel(k_\perp,k_\parallel), \quad {\mbox{where}}\\
Y(k_\perp,k_\parallel)=-j\sqrt{\frac{\varepsilon_0}{\mu_0}}\frac{2
b}{(1 - {k_\parallel^2/k_0^2})\gamma(k_\perp,k_\parallel)}.
\label{eq:tok}
\end{eqnarray}
We use time dependence of the form $\exp(+j\omega t)$. In
Eq.~(\ref{eq:tok}) $b$ is the grating period, and
$\gamma(k_\perp,k_\parallel)$ can be calculated as follows ($r_0$
is the wire radius):

\begin{equation}
\begin{array}{c}
\displaystyle \gamma(k_\perp,k_\parallel) = {k_0b\over \pi}\Bigg[\log{b\over 2\pi r_0} +\\
\displaystyle {1\over
2}\sum_{n=-\infty}^\infty\left({2\pi\over\sqrt{(2\pi n+k_\perp
b)^2 + (k_\parallel^2-k_0^2)b}}-{1\over |n|}\right)\Bigg].
\end{array}
\end{equation}

From \r{tok} it is seen that the grating susceptibility to the
incident electric field has a pole when $k_\parallel$ approaches
$\pm k_0$. Because we are interested only in the extraordinary
modes, it is suitable to express the incident electric field as
$E_\parallel = Z^e H_\perp$, where $Z^e = E_\parallel/H_\perp$ is
given by \r{fsimp} (with plus sign). Then, we get for the wire
current: \e I_{\rm w}(k_\perp,k_\parallel) = {2k_0b
H_\perp(k_\perp,
k_\parallel)\over\gamma(k_\perp,k_\parallel)\sqrt{k_\perp^2 +
k_\parallel^2 - k_0^2}}. \f This expression remains finite at
$k_\parallel = \pm k_0$.

It is of practical interest to find the field produced by the
induced grating currents. Due to the discrete nature of the
grating, it will produce all possible Floquet harmonics with
spatial frequencies $2\pi n /b$, $n\in Z$. However, the most
interesting is the case of a dense grid: $|k_\perp| < 2\pi/b$. In
this region we may consider only the main Floquet mode. The main
Floquet mode of the scattered magnetic field at the grating plane
can be found by the averaged boundary condition at the grating
plane: $H^s_\perp = \pm{I_w/(2b)}$, therefore \e
H^s_\perp(k_\perp,k_\parallel) = \pm {k_0 H_\perp(k_\perp,
k_\parallel)\over\gamma(k_\perp,k_\parallel)\sqrt{k_\perp^2 +
k_\parallel^2 - k_0^2}}. \f Here plus corresponds to the
illuminated side of the grating and minus corresponds to the
opposite side. The transmission coefficient trough the grating
(for the main Floquet mode) reads \e T(k_\perp, k_\parallel) = 1
-{k_0 \over\gamma(k_\perp,k_\parallel)\sqrt{k_\perp^2 +
k_\parallel^2 - k_0^2}}. \f

\begin{figure}[h]
\centering
\includegraphics[width=0.6\textwidth]{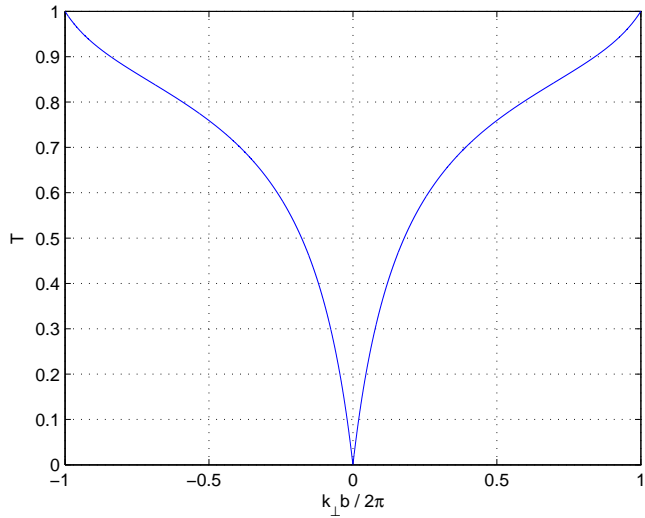}
\caption{The absolute value of the transmission coefficient for
the main Floquet mode as a function of the normalized spatial
frequency $k_\perp b/(2\pi)$. $k_\parallel=\pm k_0$ in this
example.} \label{trans}
\end{figure}

The plot of the transmission coefficient as a function of
$k_\perp$ is given in Fig.~\ref{trans}.

It is seen that the grating is less transparent to the spatial
harmonics with lower spatial frequencies. From here it becomes
clear that the improving of the resolution in the presence of the
wire grating can be explained as filtering out unwanted
low-frequency components of the spatial spectrum.
We have experimentally checked if the effect is due to possible excitation of transmission-line
modes in {\em finite} gratings by bringing absorbers to both
ends of the grating. The image was practically not changed, which
proves that the effect is indeed due to filtering of lower
spatial harmonics.

It should be mentioned that the set-up used in the experiment is
tuned to the case given in Fig.~\ref{trans}. Indeed, in the near
field of two $\lambda/2$ dipoles oriented along the wires of the
grid the spatial harmonics with $k_\parallel = \pm k_0$ dominate.
In the orthogonal direction the dipoles are seen almost as point
objects, so that the $k_\perp$ spectrum is wide and it contains
components with high spatial frequency which are not blocked by
the grating.

\section{Conclusions}

We have demonstrated, both theoretically and experimentally, that
the resolution of a near-field imaging system can be enhanced by
positioning a dense grating of thin parallel conductors before the
image plane. The effect is due to filtering out the field spectral
components with low spatial frequencies.

We have also shown that a direct comparison of silver films (at
optical frequencies) and wire gratings (at microwaves) is not
possible because of the spatial dispersion in wire media. One can
note that the physical phenomena in the considered system of wires
that lead to the resolution enhancement are rather different than
in silver films, although silver films and wire media can be both
modeled by negative effective permittivity. In wire gratings, the
effect is simply due to increasing of the grating transparency for
the spectral components of high spatial frequencies, and not due
to excitation of surface plasmons as in the case of silver films.

\end{document}